%% file: Main.tex
\documentclass[a4paper,12pt]{article}

\def\disp#1{$\displaystyle #1$}

\usepackage{graphicx}
\usepackage{enumerate}
\usepackage{rsfs}
\usepackage{amsmath}
\usepackage{amssymb}
\newcounter{subfigure}[subsection]

\title{On the Dominance of Trivial Knots among SAPs\\ on a Cubic Lattice}
\author{$^1$Akihisa Yao, $^1$Hiroshi Matsuda, $^2$Hiroshi Tsukahara,\\
$^3$Miyuki K. Shimamura and $^4$Tetsuo Deguchi\\
\vspace{3mm}\\
$^1$Department of Physics, Faculty of Science and Engineering,\\
Chuo University\\
1-13-27 Kasuga, Bunkyo-ku, Tokyo 112-8551, Japan\\
\\
$^2$Geographic Infomation Systems Department,\\
Hitachi Software Engineering Co., Ltd.\\
Kita-3, Nishi-3, Chuo-ku, Sapporo 060-0003, Japan\\
\\
$^3$Graduate School of Frontier Siences,\\
University of Tokyo\\
7-3-1 Hongo, Bunkyo-ku, Tokyo 113-8656, Japan\\
\\
$^4$Department of Physics, Faculty of Science\\
and Graduate School of Humanities and Sciences,\\
Ochanomizu University\\
2-1-1 Ohtsuka, Bunkyo-ku, Tokyo 112-8610, Japan}
\date{19 July 2001}
\begin{document}
\maketitle
\abstract
\input{Abstract}

\baselineskip=8mm

\input{Section1}

\input{Section2}
\input{Section3}
\input{Section4}

\input{Append.tex}

\input{Ref.tex}

\section*{Figure Caption}

\input{Caption}

\end{document}

%% file: Abstract.tex

The knotting probability is defined by the probability with which an $N$-step self-avoiding polygon (SAP) with a fixed type of knot appears in the configuration space.
We evaluate these probabilities for some knot types on a simple cubic lattice.
For the trivial knot, we find that the knotting probability decays much slower for the SAP on the cubic lattice
than for continuum models of the SAP as a function of $N$.
In particular the characteristic length of the trivial knot that corresponds to a `half-life' of the knotting probability is estimated to be $2.5 \times 10^5$ on the cubic lattice.

%% file: Section1.tex

\section{Introduction}
\setcounter{equation}{0}
\indent

The Self-Avoiding Polygon (SAP) with fixed topology 
gives a simplified model of real ring polymers  in solution that have 
 a topological constraint as well as  excluded volume.
Throughout the time evolution,   
a circular polymer  keeps the same knot which is given to it when it is made;  
it does not change its topology under any thermal fluctuations 
since no crossing through itself is allowed.  
On the other hand,   SAP corresponds to  
the special case of the Self-Avoiding Walk (SAW) that returns to the origin.  
If we construct a set of SAPs, then 
their topological states may contain several different knots. 
Therefore, it is not trivial how to realize the topological constraint on an SAP.  
One possible method for assigning the topological constraint on an SAP
is that after generating a large number of SAPs we 
select only such  SAPs that have the same given knot.    
By this method, we can assign the topological constraint on any model of SAP.     
In this context, the probability that a given SAP 
has a fixed knot plays a central role,    
and we call it the {\itshape knotting  probability} of the model of SAP for the  knot.
Among many different models, 
the SAP  on the cubic lattice with fixed topology     
is one of the most fundamental models of SAP.  
It has an advantage that  the  definition is very simple. We expect that   
the model should be suitable for general and mathematical study. 
In fact,  several rigorous results on knotting probability have been derived 
for the SAP on the cubic lattice \cite{Sumners,Pippenger,Whittington,Soteros}.
Thus, the  main motivation of the present research is to characterize the 
 knotting probability of the SAP on the cubic lattice 
through numerical simulations.

\par 
Let us discuss some previous numerical results  on knotting probabilities of SAP 
\cite{Vologodskii74,Wiegel,LeBret,Chen,Klenin,Koniaris,Rensburg,Deguchi94,Deguchi95,Deguchi97,Deguchi96,Orlandini}.
 For a model of SAP (or random polygon)  with $N$ steps,  
we denote by \disp{P_K(N)} the knotting probability of the model for knot $K$.  
For the Gaussian model of random polygon and the rod-bead model of SAP, 
 knotting probabilities   were evaluated 
through numerical simulations  for the trivial knot $K=\emptyset$
 \cite{Wiegel,Chen,Koniaris} and also for some non-trivial knots 
 \cite{Vologodskii74,Deguchi94,Deguchi95,Deguchi97,Deguchi96}. 
It was found that for the models   
 of SAP and random polygon,  the knotting probability 
 as a function of the step number $N$  
is given by  the following: 
\begin{equation} 
P_K(N) = C(K)N^{m(K)} \exp[-N/N(K)],
\label{formula}
\end{equation} 
where $C(K)$, $m(K)$ and $N(K)$ 
are fitting parameters. 
For large $N$, the formula (\ref{formula}) is consistent with 
the asymptotic expansion of the partition function of 
the SAP with the fixed knot type. For finite $N$, although it is not rigorous,  
it seems that the formula (\ref{formula}) fits numerical data well.  
%
%
%
 From the numerical result  it was  conjectured that the parameter  $N(K)$, 
which we call  the characteristic length of knot $K$, 
should be given by the same value for any knot $K$ \cite{Deguchi94}.
 Furthermore, it was also conjectured that the parameter $m(K)$, which we call 
the exponent of knot $K$, should be universal for different models of SAP 
or random polygon \cite{Deguchi95,Deguchi97,Deguchi96}.
We note that  the fitting formula (\ref{formula}) together with 
the two conjectures are consistent with the standard 
asymptotic behavior expected for SAP or random polygon. 
Here we also note that the rod-bead model is an off-lattice model of SAP. 

\par 
Recently, the SAP on the cubic lattice with fixed knot was studied  
through a numerical simulation using the BFACF algorithm,  
which  generates  SAPs with the same fixed knot 
but different step numbers $N$ \cite{Orlandini}.
In the simulation, the exponent $m(K)$ and the growth constant 
for the number of allowed configurations of the SAP with knot $K$ has been estimated 
for some knots. 
Furthermore, it was shown that the knotting probability 
for the trivial knot decays ``exponentially"  for a face-centered cubic lattice 
\cite{Rensburg} and for the cubic lattice \cite{Whittington}.
However, any precise estimate of the knotting probability
or the characteristic length $N(K)$ has not been given for the cubic lattice. 
Thus, it is the primary purpose of this paper 
to evaluate the characteristic length $N(K)$ for the SAP on the cubic lattice.

\par 
In 1962, Delbr\"uck \cite{Delbruck} noticed  
that the topological constraint may be  very important 
 for polymers in biology and chemistry.
Since then,  the topological problem has been studied 
in several approaches in physics and  mathematics. 
As one of the earliest  studies, des \,Cloizeaux and Mehta \cite{des Cloizeaux} estimated through numerical simulations
the critical exponent $\nu$ for the internal distance of the Gaussian random polygon 
 and discussed  some possible properties of random polygons under the topological constraint.

\par 
After the rediscovery of the pivot algorithm, 
many properties of SAW and SAP have been investigated 
not only in the field theory but also by  computer simulations \cite{Madras,MOS pivot}.
It seems, however, that there are only a few works such as  \cite{Orlandini,Tesi} 
where  the gyration radius $\langle R^2_G \rangle_{SAP}$ of the SAP on the cubic lattice 
is studied by numerical simulations.  
Let us denote by  $\langle R^2_G \rangle_{SAW}$ and $\langle R^2_G \rangle_{SAP}$ 
 the mean square  of  the gyration radius  of SAW and SAP, respectively. 
It is interesting to evaluate  
the universal amplitude ratio between $\langle R^2_G \rangle_{SAP}$ and $\langle R^2_G \rangle_{SAW}$.  
The ratio  has been  evaluated only   
up to $O(\varepsilon)$ in the $\varepsilon$-expansion method \cite{Prentis}.
Furthermore, according to the scaling theory of polymers, 
the exponent $\nu_{\sf SAP}$ for $\langle R^2_G \rangle_{SAP}$  
should be given by the exponent $\nu_{\sf SAW}$ for  $\langle R^2_G \rangle_{SAW}$. 
The agreement  is confirmed up to $O(\varepsilon)$ by the renormalization group theory \cite{Prentis,Lipkin}. 
However, it is nontrivial  to confirm the agreement  for the SAP 
on the cubic lattice through numerical simulations. 
Thus, the numerical study of the gyration radius  for the SAP on the cubic lattice 
is another purpose of this research.

\par 
Hereafter in this introduction, 
we explain some of the main results of  our numerical simulations.  
Employing the pivot algorithm, we construct a large number of SAPs 
for the SAP on the cubic lattice with given step number $N$.     
For  the gyration radius, we have obtained the exponent: $\nu_{\sf SAP} = 0.5867 \pm 0.0017$.   
This is indeed in good agreement with the estimate of the critical exponent of the SAW 
in the $\varepsilon$-expansion:
$\nu_{\sf SAW} = 0.5882 \pm 0.0011$.
Thus, our simulation in the 
paper  also confirms the agreement of the two exponents.

\par 
Let us explicitly consider the method for the topological constraint 
on the SAP in the cubic lattice with given step number $N$. 
The pivot algorithm of the SAP can 
generate all the allowed configurations of the SAP with equal probability. 
Therefore, the set of SAPs generated by the algorithm may contain 
various knots. Suppose that we have constructed $M$ SAPs of step number $N$.  
Calculating some knot invariants for each of the SAPs, we effectively  detect  
the knot types of the SAPs.
We enumerate the number of such SAPs that have the same set of 
values of the knot invariants for knot $K$,  and denote it by $M(K)$. 
The expectation value of a physical quantity 
under the topological constraint with fixed knot $K$
can be effectively calculated by taking the 
statistical average of the quantity for the $M(K)$ SAPs. 

\par 
We now turn to the knotting probability. 
Let the symbol \disp{P_K(N)} denote the knotting probability of the SAP 
with $N$ steps on the cubic lattice for knot $K$.  
If there are $M(K)$ SAPs in  the total $M$ SAPs, then 
we evaluate it  by \disp{P_K(N)}$=M(K)/M$.
In our simulation, 
 $10^5$ SAPs ($M=10^5$) are constructed  
for six different values of the step number from $N=500$ to $N=3000$. 
We have found that 
almost all SAPs are topologically equivalent to the trivial knot, 
and also that the resulting values of the knotting probability for  the trivial knot are 
fitted well by the formula (\ref{formula}). 
Thus, for the trivial knot, 
we have obtained an estimate of the characteristic length
\begin{equation} 
N(\emptyset) = (2.5 \pm 0.3) \times 10^5. 
\label{N0} 
\end{equation} 
This result means that 
trivial knots are dominant among SAPs on the cubic lattice when  
the step number $N$ is less than $10^5$.
It implies that when $N >  N(\emptyset)$, the majority of SAPs on the cubic lattice 
 have some non-trivial knots.  
The large value of the characteristic length  might be 
a consequence of strong self-avoiding effect 
of the SAP on the cubic lattice.

\par 
This paper is organized as follows.
In section 2 we will estimate the universal amplitude ratio 
and the critical exponent $\nu_{\sf SAP}$ of the gyration radius.
In section 3 we will explicitly discuss the knotting probability for the trivial knot, 
and obtain the estimate of the characteristic length.
We will also show that the knotting probability for the trivial knot 
decays almost linearly since the characteristic length of the trivial knot is so large.
Finally, we will discuss a possibility that the characteristic lengths 
take an unique value without depending on knot types, in the last section.

%% file: Section2.tex

\section{The mean-square of the gyration radius}
\setcounter{equation}{0}


\subsection{Previous results on SAW and SAP}

\indent

An $N$-step Self-Avoiding Walk (SAW) $w$ in \disp{\mathbb{Z}^3} is a sequence \disp{w_0, w_1,\cdots,w_N} of $N+1$ distinct points in \disp{\mathbb{Z}^3} such that each point $w_i$ is one of the nearest neighbors of its predecessor $w_{i-1}$: \disp{|w_i - w_{i-1}|=1 \ \textrm{\upshape for}\ i =1,\cdots,N}.
It is also subject to a constraint that any site never be occupied by two or more points in a sequence $\{w_i\}$.
The points \disp{w_0} and \disp{w_N} are the endpoints of $w$.
The components of \disp{w_i} are represented by \disp{w_i^{(\alpha)}\ \textrm{\upshape for}\ \alpha=1,2,3}.
The Self-Avoiding Polygon (SAP) is a special case of the SAW that makes a ring.
We consider an ($N-1$)-step SAW and denote it by \disp{w=\{w_0, w_1,\cdots,w_{N-1}\}}.
If the endpoint $w_0$ is the nearest neighbor of the endpoint $w_{N-1}$, this is an $N$-step SAP in \disp{\mathbb{Z}^3}.

\par 
The mean-square of the gyration radius of the SAP $\langle R^2_G \rangle_{SAP}$ 
is smaller than that of the SAW $\langle R^2_G \rangle_{SAW}$.
This fact comes from the following difference.
The endpoints of SAW are free, while those of SAP are constrained, that is, 
the endpoints of SAP should meet at the nearest neighbor sites of the lattice.
This means that the SAW has a possibility of a longest end-to-end distance whereas the SAP does not.

\par 
The RG argument gives several results of the SAW and the SAP.
Among these, we are most interested in the amplitude ratio 
$\langle R_G^2 \rangle_{SAP}/\langle R_G^2 \rangle_{SAW}$.
According to the RG theory, the amplitude ratio should be universal, 
{\it i.e.} it does not depend on the details of the models for the SAW or the SAP. 
The ratio has been evaluated by using the RG equation 
and the cluster expansion up to $O(\varepsilon)$ \cite{Prentis,Lipkin}:
\begin{equation}
{\langle R_G^2 \rangle_{SAP} \over \langle R_G^2 \rangle_{SAW}} = 0.568.\label{amp ratio rg}
\end{equation}
The ratio has also been estimated in Ref. 
\cite{Tesi} through numerical simulations with $N \le 800$: 
It is given by $0.538 \pm 0.006$.  

\par 
Another interesting result of the RG argument is that 
the critical exponent for the mean-square of the gyration radius is universal. 
It is well-known that the exponent $\nu_{\sf SAW}$ of SAW corresponds to the critical exponent 
of the {\rmfamily O}$(n)$ vector model in the limit of $n$ going to zero. 
The precise estimate of the critical exponent $\nu_{\sf SAW}$  has been made 
by the $\varepsilon$-expansion method through this correspondence
\cite{Zinn-Justin80,Zinn-Justin85,Zinn-Justin89,Zinn-Justin97}:
\begin{equation}
\nu_{\sf SAW} = 0.5882 \pm 0.0011. \label{nu rg}
\end{equation}
The exponent $\nu_{\sf SAW}$ has also been precisely evaluated by the Monte Carlo simulations, 
where the best estimate is given by the following \cite{Madras}:
\begin{equation}
\nu_{\sf SAW} = 0.5877 \pm 0.0006. \label{nu mc2}
\end{equation}

\subsection{The amplitude ratio}
\label{The universal amplitude ratio and the critical exponent}
\indent

In our simulation of SAPs on the cubic lattice, 
we have employed a length-conserving dynamical algorithm which 
keeps endpoints fixed. This algorithm was investigated in detail 
by Madras, Orlitsky, and Shepp \cite{Madras,MOS pivot}. 
We call this algorithm MOS pivot for short. 
Some details are described in Appendix A. 

\par 
We calculate the ratio $\langle R_G^2 \rangle_{SAP}/\langle R_G^2 \rangle_{SAW}$ 
and the exponent $\nu_{\sf SAP}$ using the MOS pivot 
and compare resulting values with theoretical and other simulated values.
The mean-square of the gyration radius is calculated by
\begin{equation}
\langle R_G^2 \rangle_{SAP} = {1 \over T}\sum_{t=1}^T\left[{1 \over N}\sum_{i=0}^N\left(w_i^{[t]}-{1 \over N}\sum_{j=0}^N w_j^{[t]}\right)^2\right], \label{gyration radius}
\end{equation}
where $T$ is the number of polygons and $N$ the step number, 
and the symbol \disp{w_i^{[t]}} denotes the $i$-th site in the $t$-th SAP in the 
sequence of SAPs given by ``200 successful MOS pivot operations''.  (See Appendix A for details.)

\par 
Let us note that the gyration radius should have the asymptotic behavior 
\begin{equation}
\langle R_G^2 \rangle_{SAP} = A_{\sf SAP} N^{2\nu_{\sf SAP}}(1+B N^{-\Delta}) \hspace{5mm} 
\textrm{\upshape as}\ N \to \infty.
\end{equation}
Choosing ${1\over2}$ for the exponent $\Delta$ of the correction term, 
the numerical data given by the equation (\ref{gyration radius}) are plotted in figure \ref{mean-square radius} 
and fitted to \disp{A_{\sf SAP} N^{2\nu_{\sf SAP}}(1+B N^{-{1\over2}})}, where $A_{\sf SAP}$, $B$ and $\nu_{\sf SAP}$ 
are fitting parameters.
Since there are not enough data to fit in a four-parameter curve in our simulation,
we have assumed the fixed value for the exponent $\Delta$.
We plot the mean-square of the gyration radius from $N=$500 to $N=$30000 in figure \ref{mean-square radius}.
Here, we set $T$, the number of polygons, to 10000.
The error bars denote one standard deviation given 
by the Poisson distribution to the number $T$ of polygons with $N$.

\par 
The three parameters are obtained by fitting the data in figure \ref{mean-square radius}
\begin{eqnarray}
\nu_{\sf SAP} &=& 0.5867 \pm 0.0017, \label{nu mc}\\
A_{\sf SAP} &=& 0.1101 \pm 0.0037, \label{amp mc}\\
B &=& -0.06 \pm 0.04.
\end{eqnarray}
The errors are subjective $68.3\%$ confidence intervals.
For the SAW, we recall that Madras {\it et al.} \cite{Madras} 
estimated the exponent $\nu_{\sf SAW}$ by the pivot algorithm up to $N=$80000. 
The estimate of $\nu_{\sf SAW}$ is given by (\ref{nu mc2}) 
together with the following estimates of 
the fitting parameters 
\begin{eqnarray}
A_{\sf SAW} &=& 0.19455 \pm 0.00007, \label{amp mc2}\\
B'&=& -0.11432 \pm 0.00465,\\
\Delta &=& 0.56 \pm 0.03.
\end{eqnarray}
Here, $B'$ corresponds to $AB$ in our estimated values. 
The estimate (\ref{nu mc}) is in good agreement with that 
of the RG argument (\ref{nu rg}) and that of the simulation for the SAW (\ref{nu mc2}).

\par 
From (\ref{amp mc}) and (\ref{amp mc2}), we obtain the amplitude ratio
\begin{equation}
{\langle R_G^2 \rangle_{SAP} \over \langle R_G^2 \rangle_{SAW}} = 0.566 \pm 0.019,\label{amp ratio mc}
\end{equation}
where an error is a subjective $68.3\%$ confidence interval.
The amplitude ratio (\ref{amp ratio mc}) is consistent 
with the estimated value of the RG argument (\ref{amp ratio rg}). 
The estimate of Ref.\cite{Tesi} is a littile smaller than (\ref{amp ratio mc}). 
However, it should be consistent with the estimate of Ref. \cite{Tesi}, 
if we consider  the fact that the step numbers $N$ of the simulations in Ref. \cite{Tesi} 
 are much less than 30000.  

%% file: Section3.tex

\section{The characteristic length $N({\emptyset})$}
\setcounter{equation}{0}
\subsection{A method of evaluation of the knotting probability}
\label{A method of evaluation of the knotting probability}
\indent

We evaluate the knotting probability in the following way.
We generate $M$ SAPs of $N$ steps and then enumerate the number $M(K)$ of those polygons
which are equivalent to a given knot type $K$.
We define the knotting probability \disp{P_K(N)} by the ratio \disp{M(K)/M}.

\par 
In our simulation, we determine knot types of SAPs 
using the 2nd-order Vassiliev-type invariant and the Alexander polynomial evaluated at $t=-1$.
The Vassiliev-type invariants have the following advantages: 
(1) we can calculate them in polynomial time, 
and (2) we can calculate them without consuming a large memory area \cite{Deguchi93}.
The Vassiliev-type invariants are not complete invariants.
However, in the practical sense we can safely say that
if the value of the Vassiliev-type invariant computed for an SAP is zero,
this SAP is a trivial knot.
We will see in section \ref{The number of unknotted polygons} that complicated knots are very rare events in our data.
Even if non-trivial knots are misidentified as the trivial knot by the Vassiliev-type invariant (this chance is very small), it would not affect the results of this paper.

\par 
For calculating the knotting probability, we generate random sequences of SAPs.
We make five seeds for each of the six step numbers $N=$500, 1000, 1500, 2000, 2500, 
and 3000, and then apply an operator $P$ on them for a large number of times (20000 times). 
Here,  $P$ denotes  200 successful MOS pivots. (See Appendix A.)
Then, we get effectively random sequences of SAPs. 
A sequence of SAPs derived from each seed has a set of 20000 SAPs. Here 
the sequence consists of 20000 $\times$ 200 successful MOS pivot operations. 
For each step number $N$, we thus get samples of 100000 SAPs. 

\par 
In order to analyze the behavior of the knotting probability, we use the fitting formula (\ref{formula}).
Here, we write it again:
\begin{equation*}
P_{K}(N) = C(K)N^{m(K)} \exp\left[-{N \over N(K)}\right],
\end{equation*}
where $C(N)$, $N(K)$, and $m(K)$ are fitting parameters.
In particular $N(K)$ is called characteristic length with knot type $K$. This formula was introduced by
Deguchi and Tsurusaki \cite{Deguchi94,Deguchi97,Deguchi96}.
They pointed out that the formula (\ref{formula}) is suitable for the knotting probabilities of the Gaussian and rod-bead models.
We will show that this is suitable also for the knotting probability $P_K(N)$ of the cubic lattice model.

\subsection{The random events of knotted polygons}
\label{The random events of knotted polygons}
\indent

Let us discuss statistics of knotted polygons generated in our simulation.
We will see in section \ref{The number of unknotted polygons} that almost all polygons are of trivial knots.
Therefore, we may assume that non-trivial knots are generated such as the Poisson random events: the number of trivial knots between two knotted SAPs will follow the Poisson distribution, if the SAPs are randomly constructed.

\par 
Let us consider the ``time interval" of the Poisson random events.
We recall that the time $t$ is a discrete number.
We measure the length $L$ of the time interval of SAPs from time $t_1$ at which a knotted SAP appears to time $t_2$ at which the next knotted SAP appears after $t_1$,
and set $L=t_2-t_1$.
If non-trivial knots are generated as the Poisson random events,
the time interval $L$ follows the function:
\begin{eqnarray}
D(N,L)=A(N)\,P_{\emptyset}(N)^{L-1}.
\end{eqnarray}
We call it the discrete distribution function of the time interval $L$.
Here, $P_\emptyset(N)$ is the knotting probability for the trivial knot.
We have introduced $A(N)$ for a technical reason and hence $D(N,L)$ is not necessarily normalized.

The discrete distribution functions of the time interval $L$ are numerically evaluated as follows.
The time interval $L$ is a discrete random variable.
We introduce a sequence of natural numbers $\{l_0,l_1,\cdots\}$, where
$l_i = 50 \times i$.
Then, we count the number of polygons with $L$ taking values in $[l_i,l_{i+1})$ and plot it at $l_i$ for each $i$. This is the discrete distribution function of the time interval $L$ obtained from numerical evaluation.

We plot the discrete distribution functions obtained by the numerical evaluation and $D(N,L)$ as a function of $L$, where $P_{\emptyset}(N)$ is estimated from table \ref{table1} and $A(N)$ is chosen to fit these distributions (See figure \ref{interval1} and \ref{interval2}).
Error bars denote one standard deviation.
They are estimated by applying the Poisson distribution to the number of samples included in interval $L$.
For $N=$1000, 1500, 2000, 2500, and 3000, 
these graphs show a fairly good agreement with the function $D(N,L)$.
In the case of N=500, the data deviate from the function.
This is not unexpected since we have too few samples of knotted polygons.
We do not plot the graph for $N=$500 in this paper.

We expect from figure \ref{interval1} and \ref{interval2} that the MOS pivot makes a uniform ensemble for knots.
Thus, we can calculate the knotting probability and the characteristic length using the MOS pivot.

\subsection{The number of unknotted polygons}
\label{The number of unknotted polygons}

\indent

The table \ref{table1} gives the number of each knot type with respect to $N$.
Here, errors correspond to 68.3\% confidence intervals.
They are estimated by applying the binomial distribution to the number $M(K)$ of polygons for knot $K$.
In the table \ref{table1} we explain the notations: $\emptyset$,$3_1$,$4_1$ denote the trivial knot, the trefoil knot and the figure eight knot, respectively.
The other knot types are denoted by $etc$.

The table \ref{table1} tells clearly that almost all the generated SAPs are trivial knots.
$M(3_1)$ is also much larger than $M(4_1)$.
The other knots ($etc.$) are nearly equal to zero.

\begin{table}[hbtp]
\begin{center}
\begin{tabular}{|r|r|r|r|r|r|}
\hline
\multicolumn{1}{|c|}{\textbf{Step Numbers}} & \multicolumn{1}{|c|}{\textbf{$M(\emptyset)$}} & \multicolumn{1}{|c|}{\textbf{$M(3_1)$}} & \multicolumn{1}{|c|}{\textbf{$M(4_1)$}} & \multicolumn{1}{|c|}{\textbf{etc.}} & \multicolumn{1}{|c|}{\textbf{$M$}}\\
\hline
 500 & 99849 $\pm$ 25 &   147 $\pm$ 24 &  3 $\pm$  3 &  1 $\pm$ 1 & 100000\\
\hline
1000 & 99640 $\pm$ 38 &   344 $\pm$ 37 &  9 $\pm$  6 &  7 $\pm$ 5 & 100000\\
\hline
1500 & 99430 $\pm$ 48 &   541 $\pm$ 46 & 24 $\pm$ 10 &  5 $\pm$ 4 & 100000\\
\hline
2000 & 99208 $\pm$ 56 &   752 $\pm$ 55 & 27 $\pm$ 10 & 13 $\pm$ 7 & 100000\\
\hline
2500 & 98965 $\pm$ 64 &   985 $\pm$ 62 & 38 $\pm$ 12 & 12 $\pm$ 7 & 100000\\
\hline
3000 & 98787 $\pm$ 69 &  1157 $\pm$ 68 & 40 $\pm$ 13 & 16 $\pm$ 8 & 100000\\
\hline
\end{tabular}
\end{center}
\vspace{-8mm}
\caption{the Number of Generated Knots}
\vspace{-4mm}
\center{The estimated errors correspond to one standard deviation.}
\label{table1}
\end{table}

Next we focus on the knotting probability for the trivial knot, and plot $P_{\emptyset}(N)$ as a function of $N$ (figure \ref{unknotting pro fig}).
The error bars are one standard deviation.
$P_{\emptyset}(N)$ decays linearly with respect to $N$.
It is expected that $P_{\emptyset}(N)$ becomes decays exponentially when $N$ goes to infinity.
In \cite{Sumners,Pippenger,Whittington} it was shown that the knotting probability $P_{\emptyset}(N)$ tends to zero ``exponentially":
\begin{equation}
P_{\emptyset}(N) = A\exp[-\kappa N +o(N)], \hspace{5mm}\textrm{\upshape when}\ N \to \infty. \label{asymptotic shape}
\end{equation}
Thus, this is a natural situation.

The asymptotic shape (\ref{asymptotic shape}) is realized for the trivial knot in our case
when the fitting parameters of the formula (\ref{formula}) take the following values: $|m(\emptyset)| \ll 1$ and $C(\emptyset) \simeq 1$.
In fact, using the least-squares estimation, we find:
\begin{eqnarray}
C(\emptyset) &=& 1.0035 \pm 0.0035\\
m(\emptyset) &=& (-4.7 \pm 5.7) \times 10^{-4}\\
N(\emptyset) &=& (2.5 \pm 0.3) \times 10^5 \\
\chi^2 &=& 0.748 \\
Prob(\chi^2>0.748) &=& 0.862
\end{eqnarray}
(errors are one standard deviation).
Here, the $\chi^2$ value ({\itshape i.e.} sum of square of normalized deviations from the regression line) can serve as a criterion of good fit.
It should be distributed as $\chi^2$ with $n-3$ degrees of freedom, where $n$ is the number of data points in the fit.
$Prob(\chi^2 >0.748)$ is the probability that $\chi^2$ would exceed the observed value, in this case $86.2\%$.
This implies that the formula (\ref{formula}) is suitable.
It is remarkable that the characteristic length of the trivial knot is much larger than the value expected from the rod-bead model.

\par 
There have been a few simulation studies on several lattices.
In \cite{Rensburg} the knotting probability for the trivial knot 
was calculated on a face-centered cubic (FCC) lattice. 
It was shown that assuming the two-parameter fitting formula: 
$P_\emptyset(N) = C(\emptyset)e^{-\alpha(\emptyset)N}$, the parameters were given by 
\begin{eqnarray*}
\alpha(\emptyset) &=& (7.6 \pm 0.9) \times 10^{-6} \\
C(\emptyset) &=& 1.0011 \pm 0.003 \\
\chi^2 &=& 2.7 \\ 
Prob(\chi^2 > 2.7) &=& 0.44,
\end{eqnarray*}
where errors were one standard deviation.
Our interpretation of $\alpha(\emptyset)$ taking $7.6 \times 10^{-6}$ 
is that the characteristic length is $1.3 \times 10^5$. 
Comparing the cubic lattice with the FCC lattice, 
the characteristic length is larger on the cubic lattice than on the FCC lattice.
In \cite{Whittington} it was also shown that the exponent $\alpha(\emptyset)$ 
in the above form was $(5.7 \pm 0.5)\times 10^{-6}$ on the cubic lattice.
This corresponds to the characteristic length of $1.8 \times 10^5$.
We can expect that this is consistent with our estimated value within the error bars. 
Here we note that the connection of Ref. \cite{Orlandini} shall be discussed 
in section \ref{A consequence of the large characteristic length} 
and also note that the simulation results of Refs.
\cite{dimensions,entropic} seem to contain  
some information on the characteristic length of the SAP on the cubic lattice. 
However, we are unable to derive any appropriate estimate from them.

\par 
Let us return to our data. We estimate not only the characteristic length $N(\emptyset)\simeq 2.5\times 10^5$ 
but also the exponent of a correction term $m(\emptyset)\simeq 0$.
These parameters show that the knotting probability 
for the trivial knot serves a ``pure exponential" decay on the cubic lattice.
This is a new result for the cubic lattice model.
Thus, the simulation in the paper improves that of \cite{Whittington}.

\par 
We have several interpretations of the characteristic length $N(\emptyset)$.
Since $N(\emptyset)$ is so large, we found that the number of knotted polygons are much smaller than that of unknotted polygons in our simulation.
We expect that the knotting probability for the trivial knot decreases to about $30\%$ at $N=N(\emptyset)$.
Thus, non-trivial knots become majority of SAPs when $N$ is larger than $N(\emptyset)$.
In addition to this, we will see in section \ref{A consequence of the large characteristic length} that $N(\emptyset) \simeq N(K)$ for any knot type $K$ is due to the fact that $N(\emptyset)$ is large.

\par
In figure \ref{knotting pro fig}, we plot the knotting probability for the trefoil knot $P_{3_1}(N)$ as a function of $N$, where error bars are one standard deviation. The data points almost lie on a straight line.
Since the data of the trefoil knot are only six points in the step number, we cannot fit the knotting probability $P_{3_1}(N)$ to formula (\ref{formula}).
We leave the selection of fitting parameters for the trefoil knot or more complicated knots as future investigations.

\par
The knotting probability $P_{3_1}(N)$ may have a finite-size effect.
When we set $m(3_1)=1$ as expected from \cite{Orlandini,Deguchi97}, the formula (\ref{formula}) does not match with $P_{3_1}(N)$ in our data.
Deguchi and Tsurusaki argued that a finite-size effect appeared in the knotting probabilities in \cite{Deguchi97}.
When we plot the straight line fitted to $P_{3_1}(N)$, the line intersects with the $x$-axis at a positive value.
This is the finite-size effect.
We introduce offset parameter $N_{ini}$ and replace $N$ by $\tilde{N}=N-N_{ini}$ in the formula (\ref{formula}).
When we fix $m(3_1)=1$, our rough estimation gives as $N_{ini} \sim 140$ for the trefoil knot.
On the other hand, we roughly estimate $N_{ini} \sim 0$ for the trivial knot.
We expect that such finite-size effect would also appear in the knotting probabilities for the SAPs with more complicated knot types.
It could be confirmed by generating SAPs with much larger $N$.

\par 
In section \ref{The universal amplitude ratio and the critical exponent}
we calculated the gyration radius including all possible knots,
and then estimated the universal amplitude ratio and the universal exponent $\nu$.
Without classifying knot types, however, we can effectively consider only trivial knots when $N < N(\emptyset)$.
The universal amplitude ratio and the universal exponent are evaluated effectively for the trivial knots,
although 2\% of the SAPs are non-trivial knots and we neglect their influence.

%% file: Section4.tex

\section{A consequence of\\ the large characteristic length $N({\emptyset})$}
\label{A consequence of the large characteristic length}
\setcounter{equation}{0}
\indent

The trivial knot dominates among SAPs on the cubic lattice when $N$ is less than $2.5 \times 10^5$.
Our interpretation is that the cubic lattice has so strongly the excluded-volume effect that it almost prevents appearances of knotted polygons.

We have a conjecture that the appearance of a complicated knot is a rare event on the cubic lattice.
According to \cite{Deguchi94,Deguchi97,Deguchi96}, non-trivial knots occupy a large number of configurations of SAPs for $N \ge N(\emptyset)$.
We believe that the above situation is realized also on the cubic lattice.
If the formula (\ref{formula}) is a suitable form of the knotting probability and if the trefoil and figure eight knots on the cubic lattice behave like these of continuum models,
the ratio $M(4_1)/M(3_1)$ for each step number $N$ should depend only on the ratio $C(4_1)/C(3_1)$.
We recall that $M(3_1) \gg M(4_1)$ for $N <$ 3000 from the table \ref{table1}.
Then, at $N \sim 2.5 \times 10^5$ almost all SAPs are expected to be $3_1$ knotted polygons on the cubic lattice unlike SAPs on continuum models.

We see that our estimated value $N(\emptyset)$ is related to the growth constant $\mu$ from the viewpoint of
Orlandini {\itshape et al.}\cite{Orlandini}.
The asymptotic behavior of the number of $N$-step polygons $c_N$ is given by
\begin{equation}
c_N = aN^{\alpha-3}\mu^N\left(1+bN^{-\Delta}+o(N^{-1})\right), \label{c_N}
\end{equation}
where $\alpha$ and $\Delta$ are critical exponents.
For fixed knot type $K$, it is believed that the number $c_N(K)$ of polygons with knot $K$ should have a similar form:
\begin{equation}
c_N(K) = a(K)N^{\alpha(K)-3}\mu(K)^N\left(1+b(K)N^{-\Delta(K)}+o(N^{-1})\right). \label{c_N(K)}
\end{equation}
Then, the knotting probability $P_K(N)$ is given by $c_N(K)/c_N$, and this implies that the characteristic length $N(\emptyset)$ relates to the growth constants $\mu$ and $\mu(\emptyset)$. We can estimate the ratio $\mu/\mu(\emptyset)$ from the value of $N(\emptyset)$:
\begin{equation}
{\mu \over \mu(\emptyset)} \simeq e^{1/N(\emptyset)} \sim 1 + (4 \pm 2)\times 10^{-6}, \label{diff} 
\end{equation}
where an error is two standard deviations.

Let us discuss the independence of the characteristic length $N(K)$ with respect to knot type $K$.
Orlandini {\itshape et al.} \cite{Orlandini} calculated the growth constants directly and showed the following equality:
\begin{equation}
\mu(\emptyset) = \mu(K) = 4.6836 \pm 0.0038 \hspace{5mm} \textrm{\upshape for}\ any\ knot\ type\ K, \label{mu K}
\end{equation}
where an error corresponds to $95\%$ confidence interval.
In addition Guttmann estimated the growth constant \cite{Guttmann}
\begin{equation}
\mu = 4.68393 \pm 0.00002 \label{mu}
\end{equation}
using exact enumeration and series analysis (an error is one standard deviation). From (\ref{mu K}) and (\ref{mu}), we obtain the ratio
\begin{equation}
{\mu \over \mu(\emptyset)} = {\mu \over \mu(K)} \simeq 1 + (7 \pm 8) \times 10^{-5}, \label{diff2}
\end{equation}
where an error is two standard deviations.
From (\ref{diff}) and (\ref{diff2}), we expect that the characteristic length is independent of the knot type: $N(\emptyset) \simeq N(K) \sim 2.5 \times 10^5$ for any knot type $K$.

Although the independence of $\mu(K)$ with respect to knot type $K$ is pointed out by
Orlandini {\itshape et al.} \cite{Orlandini}, we can also confirm it more precisely through our simulation.
While Orlandini {\itshape et al.} \cite{Orlandini} calculated the growth constants, we have no direct calculation of them.
However, we can predict that the differnce between $\mu(\emptyset)$ and $\mu(K)$ is very small when we use the following inequalities: \disp{\liminf_{N \to \infty}N^{-1}\log c_N(K) \ge \mu(\emptyset)} and \disp{\limsup_{N \to \infty}N^{-1}\log c_N(K) < \mu} for any knot type $K$, which were proven in \cite{Soteros,Whittington}.
If both \disp{\liminf_{N \to \infty}N^{-1}\log c_N(K)} and \disp{\limsup_{N \to \infty}N^{-1}\log c_N(K)} exist
and take the same value $\mu(K)$,
then we have $\mu(\emptyset) \le \mu(K) < \mu$.
These inequalities and the estimate (\ref{diff}) limit the ratio $\mu(K)/\mu(\emptyset)$ to
\begin{equation}
\left|{\mu(K) \over \mu(\emptyset)} - 1 \right| \le (4 \pm 2) \times 10^{-6}.
\end{equation}
This is a strong bound.
Thus, we can expect that $\mu(\emptyset) \simeq \mu(K)$ for any knot type $K$.

\section*{Acknowledgement}
\indent

We would like to thank Takeo Inami for a careful reading of the manuscript and valuable comments.
A.Y. is supported by a Research Assistant Fellowship of Chuo University.
This work is partially supported by the Grant-in-Aid for Encouragement of Young Scientists (No. 12740231).

%% file: Append.tex
\vspace{2mm}
\appendix
\hspace{-6mm}{\Large\bfseries Appendix}
\section{The method for generating SAPs}
\indent

Let us discuss the method for making SAPs in the paper. 
We first construct a seed SAP $w$,  and then derive a random sequence of SAPs 
 $\{w^{[1]},w^{[2]},\cdots,w^{[t-1]},w^{[t]},\cdots\}$, where $w^{[t]}=P(w^{[t-1]})$ and $w^{[0]}=w(0)$.
Here the operator $P$ will be  defined later.

\par 
We construct a `seed' SAP, by combining two SAWs which have the same endpoints, in the following way: 
first we make an ${N/2}$-step SAW using the myopic self-avoiding walk (MSAW) algorithm \cite{Madras Slade}, 
where $N$ is an even integer; 
secondly we perform the MOS pivot transformations with respect to \disp{k=0,l=N/2}. 
\cite{MOS pivot};  
finally we concatenate the endpoints of the new and original SAWs respectively (figure \ref{seed SAP}), 
and get an SAP with the step number $N$ if it has no self-intersections. 

\par 
Let us discuss how to construct  a random sequence of SAPs in our simulations. 
If one of the MOS pivot transformations changes a given SAP into a different SAP, 
this operation is called a successful MOS pivot operation. 
We consider a sequence of successful MOS pivot operations. 
Then, we define an operator $P$ by 200 successful MOS pivot operations in the sequence. 
We note that the SAP obtained by a single successful MOS pivot 
is not independent from the original one: they are correlated. 
However, the correlation decays almost completely after  200 successful MOS pivot operations, which shall be 
shown in Appendix B.  Thus, we may consider that for any given SAP $w$, $P(w)$ is independent from $w$. 

\par 
For generating SAPs, we use the Mersenne Twister that is a pseudo random number generator \cite{MT}.
This algorithm has the following properties:
(1) we can get many samples because the period is $2^{19937}-1$,
(2) we treat high dimensional space (max $623$ dimensions),
(3) pseudo random numbers are generated fast and
(4) we can use the memory efficiently.
Thus, the Mersenne Twister is a high performance generator.

\section{The decay of  correlations among the SAPs obtained by the MOS pivot operations}
\indent

In order to check the validity of  the random sequence of SAPs constructed in the paper, 
we show explicitly how the correlation among the SAPs decays 
after applying a number of  MOS pivot operations.

\par 
Let us regard the number $\tau$ of successful MOS pivot operations  
as the time of evolution of the shape of SAP under the sequential MOS pivot operations: 
the seed SAP $w$  of a random sequence corresponds to $w(0)$ for $\tau=0$;  
$w(\tau)$ is defined by $p^\tau(w)$ for $\tau>0$. 
Here  $p$ have denoted a successful MOS pivot operation. 
Let us now define the correlation function for the structure of SAP with the step number $N$  
in the following 
\begin{equation}
C(\tau) = \frac{\sum_{\alpha=1}^3\sum_{i=0}^{N-1}\left\langle\left(w_i^{(\alpha)}(0)-{1\over N}\sum_{j=0}^{N-1}w_j^{(\alpha)}(0)\right)\left(w_i^{(\alpha)}(\tau)-{1\over N}\sum_{l=0}^{N-1}w_l^{(\alpha)}(\tau)\right)\right\rangle}{\sum_{\alpha=1}^3\sum_{i=0}^{N-1}\left\langle\left(w_i^{(\alpha)}(0)-{1\over N}\sum_{j=0}^{N-1}w_j^{(\alpha)}(0)\right)^2\right\rangle},\label{correlation function}
\end{equation}
where \disp{w_i^{(\alpha)}(\tau)} is the $\alpha$-component of the $i$-th site of SAP (or SAW) 
after $\tau$ pivot operations and $\langle\cdot\rangle$ denotes the statistical average.

\par 

\par 
In figure \ref{correlation1} and \ref{correlation2}, we plot the correlation $C(\tau)$ 
versus the time $\tau$ in the case of $N =$ 500 and 1000, respectively.
Error bars show one standard deviation which are estimated by assuming 
that the data follows the Poisson distribution.
In our simulation, we evaluated the correlations of SAPs 
with the step number $N$ at 500 and 1000, 
and generated 10000 sequences (starting from 10000 different seeds) to take the statistical average.

\par 
The decay rate of the correlation function $C(\tau)$ 
is slower in the case of the MOS pivot 
than in the case of normal pivot algorithm (figure \ref{correlation1} and \ref{correlation2}).
This is due to the difference in the numbers  of independent pivot transformations. 
Note that the number of independent transformations 
for the normal pivot algorithm \cite{Madras,Madras Slade} is $d!2^d\!-\!1$
(the normal pivot algorithm is an algorithm with one endpoint free while the other fixed)
, which correspond to the number of all the elements of $d$-dimensional orthogonal group on a hypercubic lattice, 
while for the MOS pivot $2d(d-1)+1$ in the $d$-dimensional hypercubic lattice.
In the cubic lattice, the MOS pivot has 13 pivot transformations while the normal pivot algorithm has 47.

%% file: Ref.tex

\bibliographystyle{plain}

%% file: Caption.tex
\setcounter{subfigure}{0}
\begin{center}
\begin{minipage}{12cm}
\rotatebox[origin=c]{-90}{
\includegraphics[width =.8\linewidth]{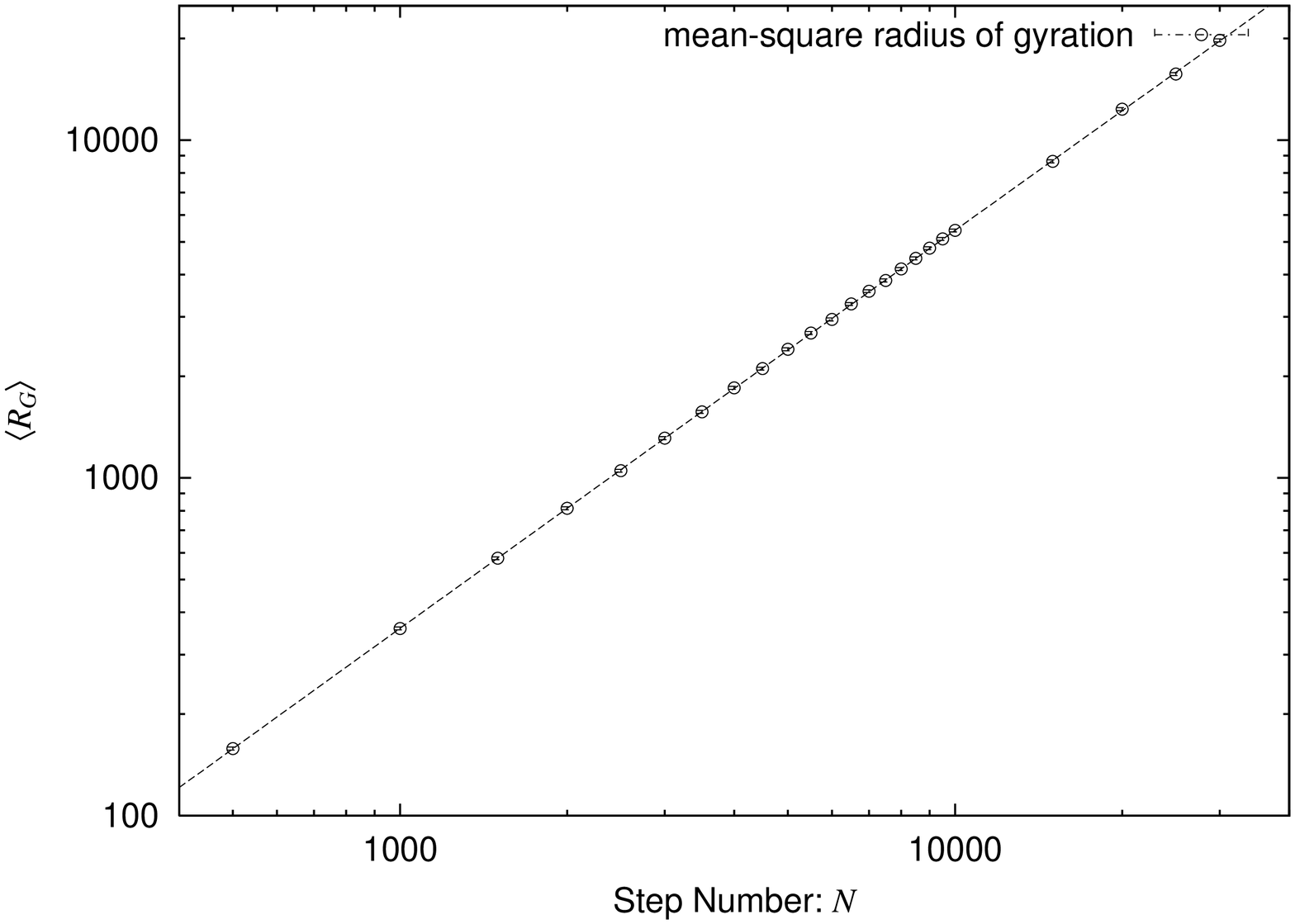}}
\end{minipage}
\end{center}
\begin{figure}[htbp]
\caption{Mean-Square Radius of Gyration $\langle R_G \rangle$}
{
We give the log-log plotted graph of the mean-square of gyration radius versus the step number $N$.
Here error bars denote one standard deviation.
}
\label{mean-square radius}
\end{figure}

\setcounter{figure}{1}
\setcounter{subfigure}{1}
\begin{figure}[htbp]
\begin{center}
\begin{minipage}{12cm}
\rotatebox[origin=c]{-90}{
\includegraphics[width =.8\linewidth]{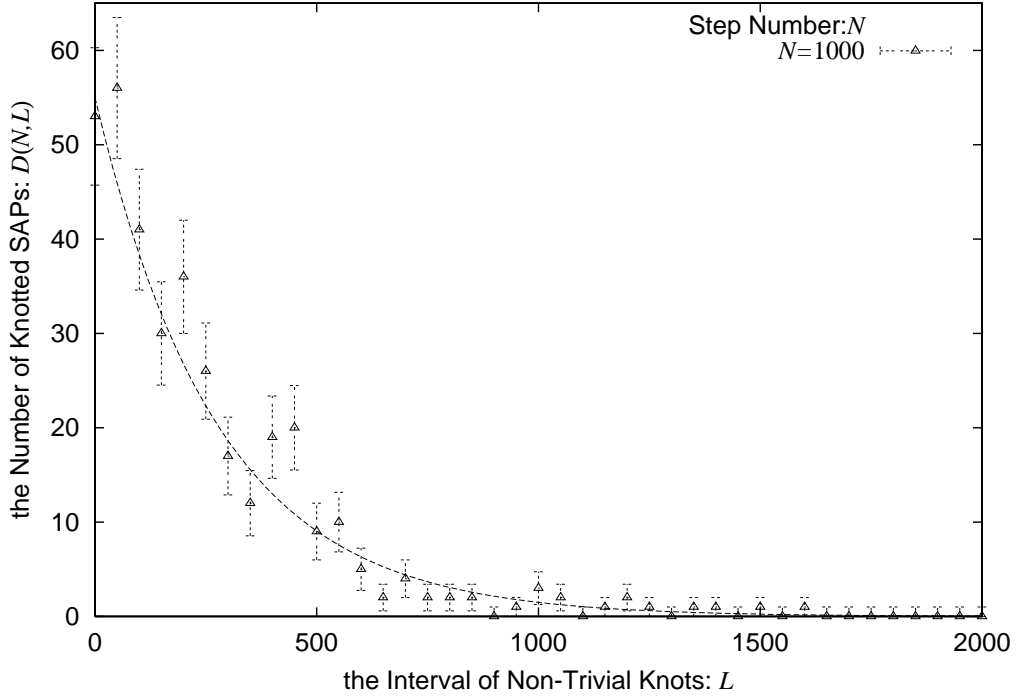}}
\end{minipage}
\end{center}
\caption{Discrete Distribution Function of Interval $L$ of Non-Trivial Knots for $N=1000$}
{
The number of occurring a knotted SAP with respect to the time interval $L$ is represented by the Poisson distribution: $D(N,L) = A(N)P_\emptyset(N)^L$.
$P_\emptyset(N)$ is the knotting probability of the trivial knot with the step number $N$ and
$A(N)$ is a normalization factor.
We call it the discrete distribution function of the time interval $L$.
In our simulation, we count the number of polygons with $L$ taking values in $[l_i,l_{i+1})$ and plot it at $l_i$ for each subscript $i$.
Here, the sequence $\{l_0,l_1,\cdots\}$ is defined by $l_i=50 \times i$.
The data are fitted to the distribution $D(N,L)$.
We measure the length of the time interval of SAPs from time $t_1$ at which a knotted SAP appears, to time $t_2$ at which the next knotted SAP appears after $t_1$, and denote it by $L$ ($ = t_2 - t_1$).
}
\label{interval1}
\end{figure}

\setcounter{figure}{1}
\setcounter{subfigure}{2}
\begin{figure}[htbp]
\begin{center}
\begin{minipage}{12cm}
\rotatebox[origin=c]{-90}{
\includegraphics[width =.8\linewidth]{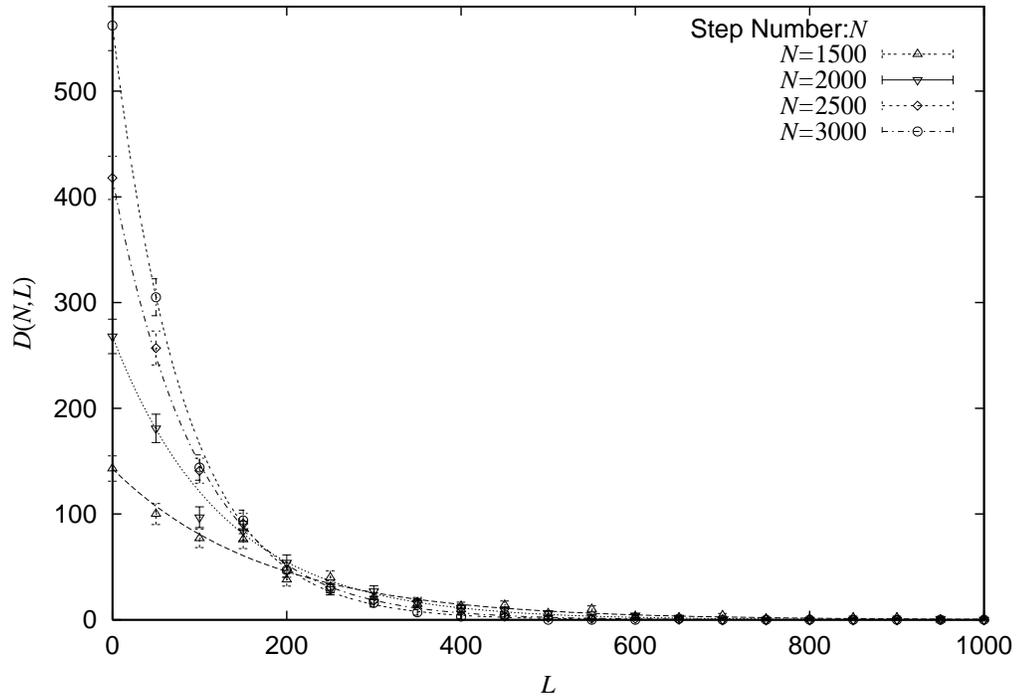}}
\end{minipage}
\end{center}
\caption{Discrete Distribution Function of Interval $L$ of Non-Trivial Knots
for $N=1500$,$2000$,$2500$ and $3000$}
{
We plot the distributions $D(N,L)$ and the data of $L$ taking values in $[l_i,l_{i+1})$ for the step number from
$N=1,\!500$ to $N=3,\!000$, respectively.
The data are fitted to $D(N,L)$ well.
Therefore, the MOS pivot generates SAPs without biasing statistics of knots.
}
\label{interval2}
\end{figure}

\setcounter{subfigure}{0}
\begin{figure}[htbp]
\begin{center}
\begin{minipage}{12cm}
\rotatebox[origin=c]{-90}{
\includegraphics[width =.8\linewidth]{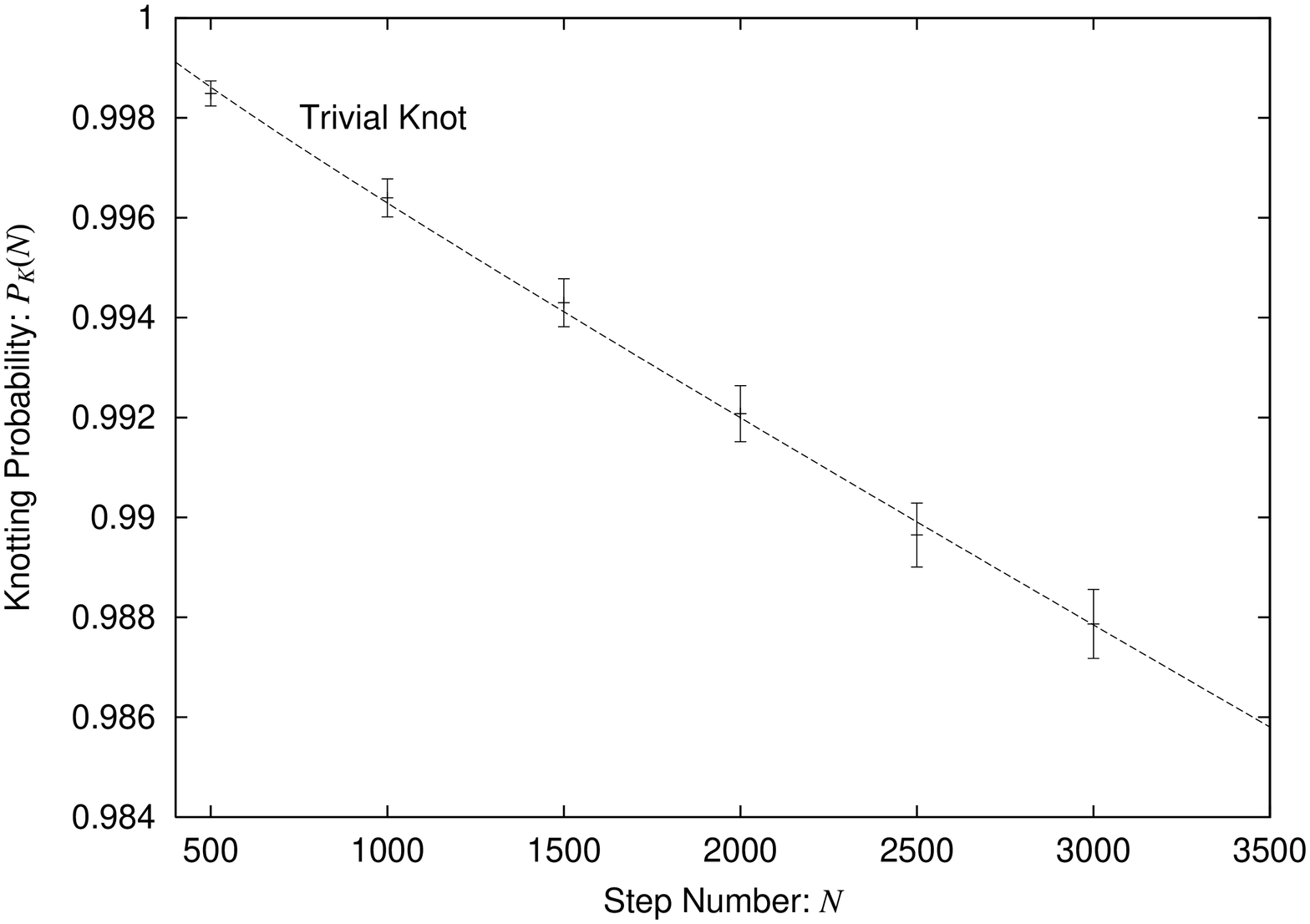}}
\end{minipage}
\end{center}
\caption{Knotting Probability $P_{\emptyset}(N)$}
{
We give the graph of the knotting probability $P_{\emptyset}(N)$ for the trivial knot versus the step number $N$.
Here error bars denote one standard deviation.
The data behaves as a liner decay with respect to $N$.
}
\label{unknotting pro fig}
\end{figure}

\begin{figure}[htbp]
\begin{center}
\begin{minipage}{12cm}
\rotatebox[origin=c]{-90}{
\includegraphics[width =.8\linewidth]{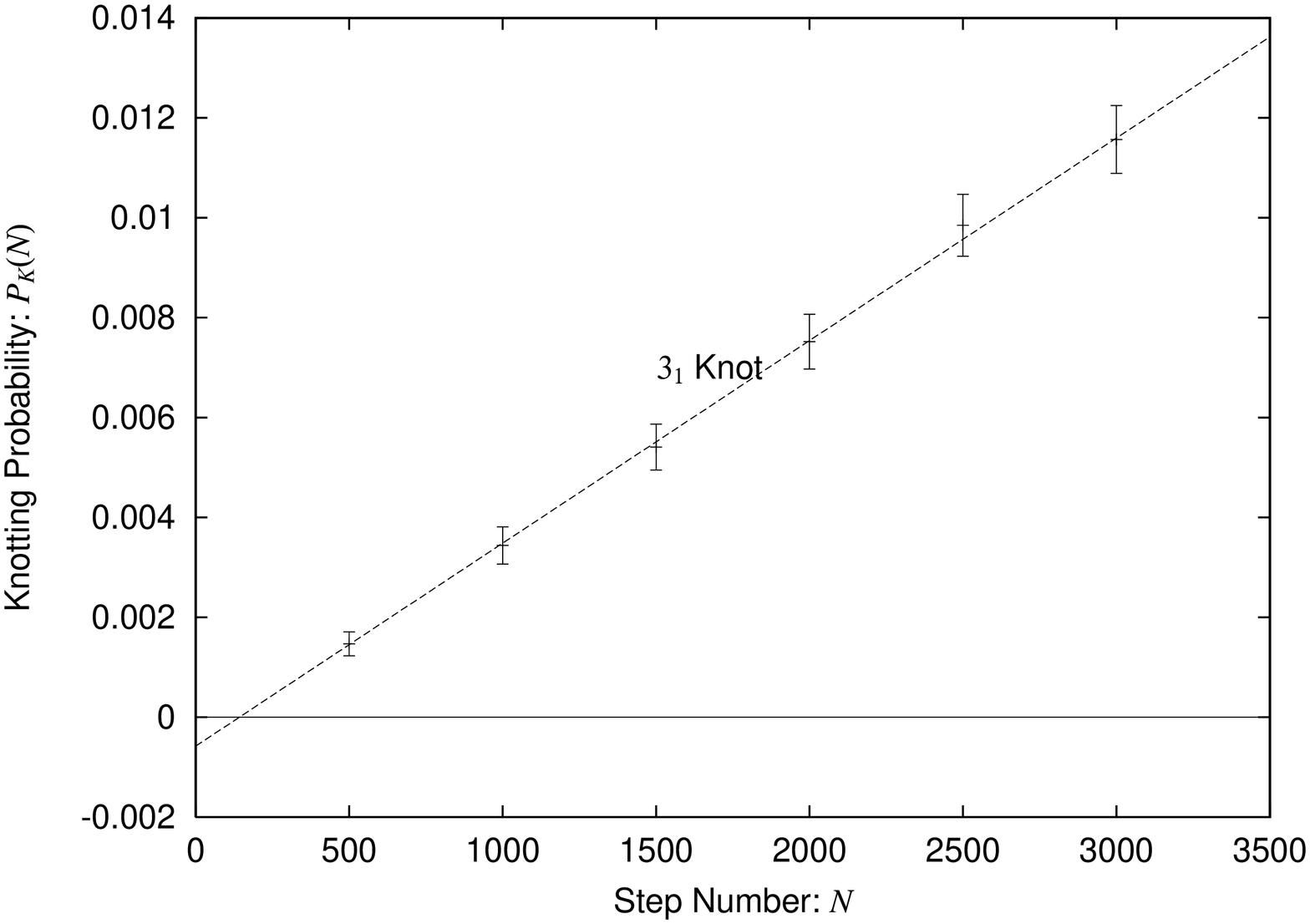}}
\end{minipage}
\end{center}
\caption{Knotting Probability $P_{3_1}(N)$}
{
We give the graph of the knotting probability $P_{3_1}(N)$ of trefoils versus the step number $N$.
Here error bars denote one standard deviation.
If we assume a finite-size effect for the SAPs of trefoils,
we fit the data to the straight line which intersect $x$-axis at positive value.
}
\label{knotting pro fig}
\end{figure}

\begin{figure}[htbp]
\begin{center}
\vspace{-80mm}
\begin{minipage}{12cm}
\hspace{-95mm}
\rotatebox[origin=c]{-90}{
\includegraphics[width =2.0\linewidth]{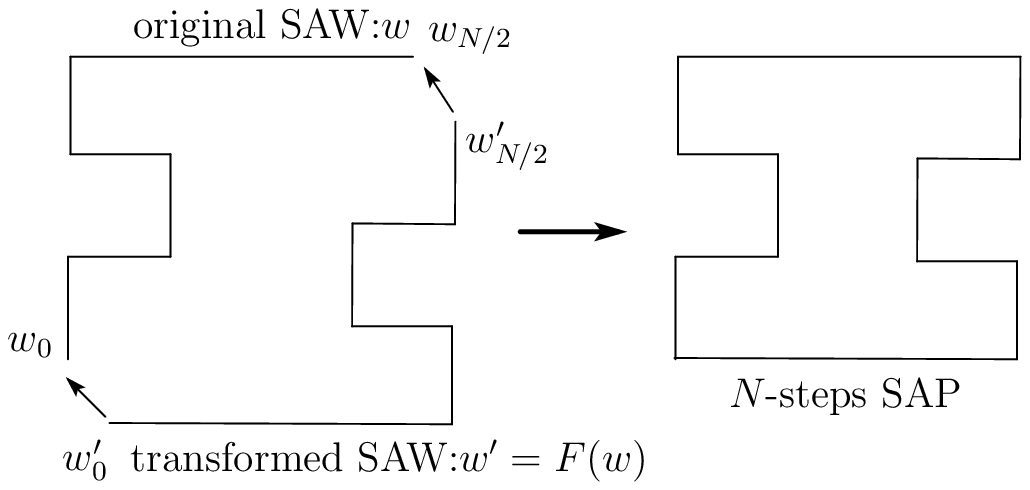}}
\end{minipage}
\end{center}
\vspace{-150mm}
\caption{Generated Seed SAP with Step Number $N$}
{
An original SAW $w$ generated by the MSAW algorithm with the step number $N/2$ transforms into another SAW using the MOS pivot with respect to $k=0$ and $l=N/2$.
We concatenate the edges of new and original SAWs, respectively. We get an $N$-step SAP.
}
\label{seed SAP}
\end{figure}

\setcounter{figure}{5}
\setcounter{subfigure}{1}
\begin{figure}[htbp]
\begin{center}
\begin{minipage}{12cm}
\rotatebox[origin=c]{-90}{
\includegraphics[width =.8\linewidth]{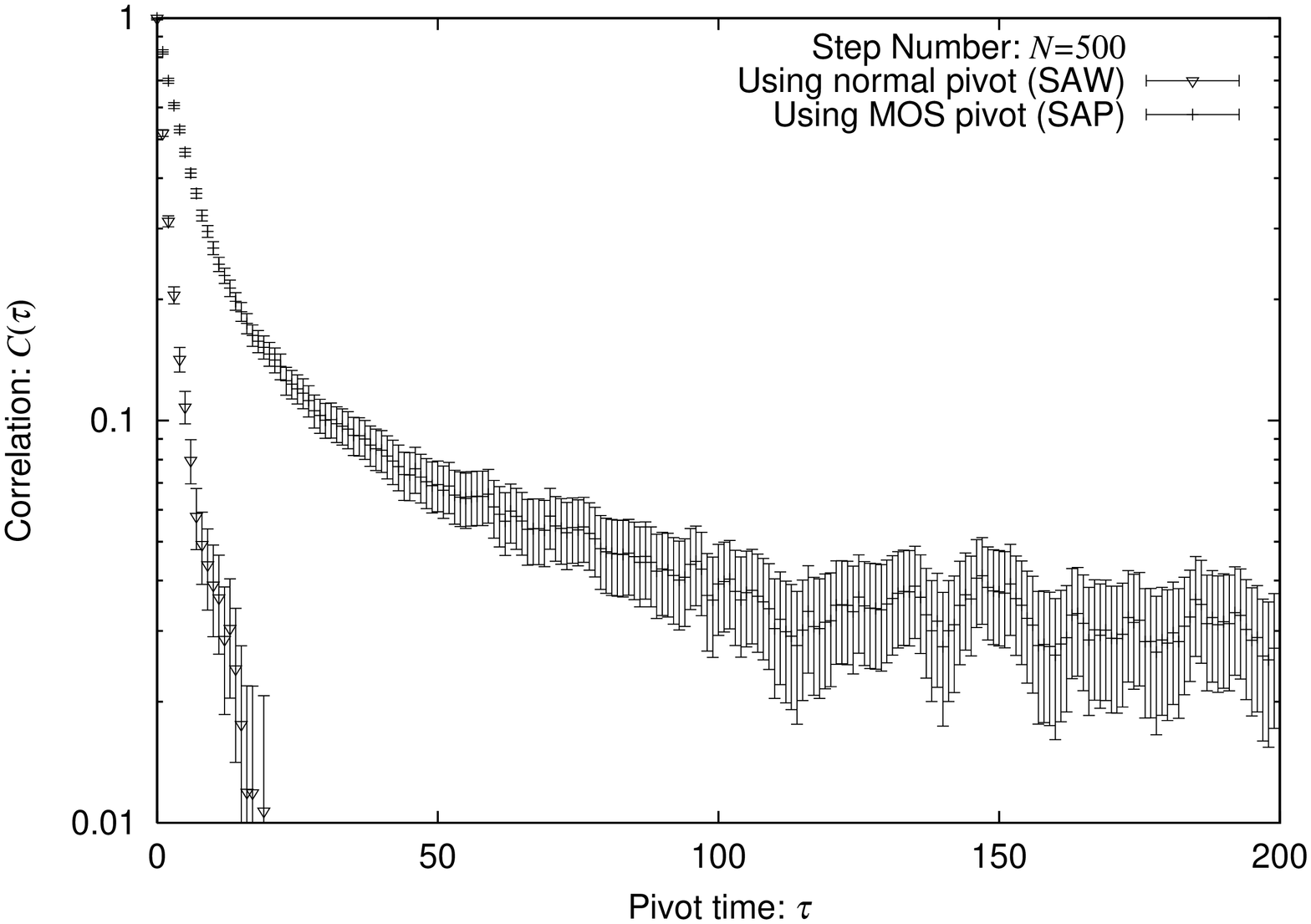}}
\end{minipage}
\end{center}
\caption{Pivot Correlation $C(\tau)$ at $N=500$}
{
We plot the pivot correlations $C(\tau)$ versus the time $\tau$.
The MOS pivot and the normal pivot operations in the case of $N=500$, respectively.
Here, the $\tau$ is the number of trials, and
error bars correspond to one standard deviation.
The correlation of the MOS pivot decays slower than that of the normal pivot.
}
\label{correlation1}
\end{figure}

\setcounter{figure}{5}
\setcounter{subfigure}{2}
\begin{figure}[htbp]
\begin{center}
\begin{minipage}{12cm}
\rotatebox[origin=c]{-90}{
\includegraphics[width =.8\linewidth]{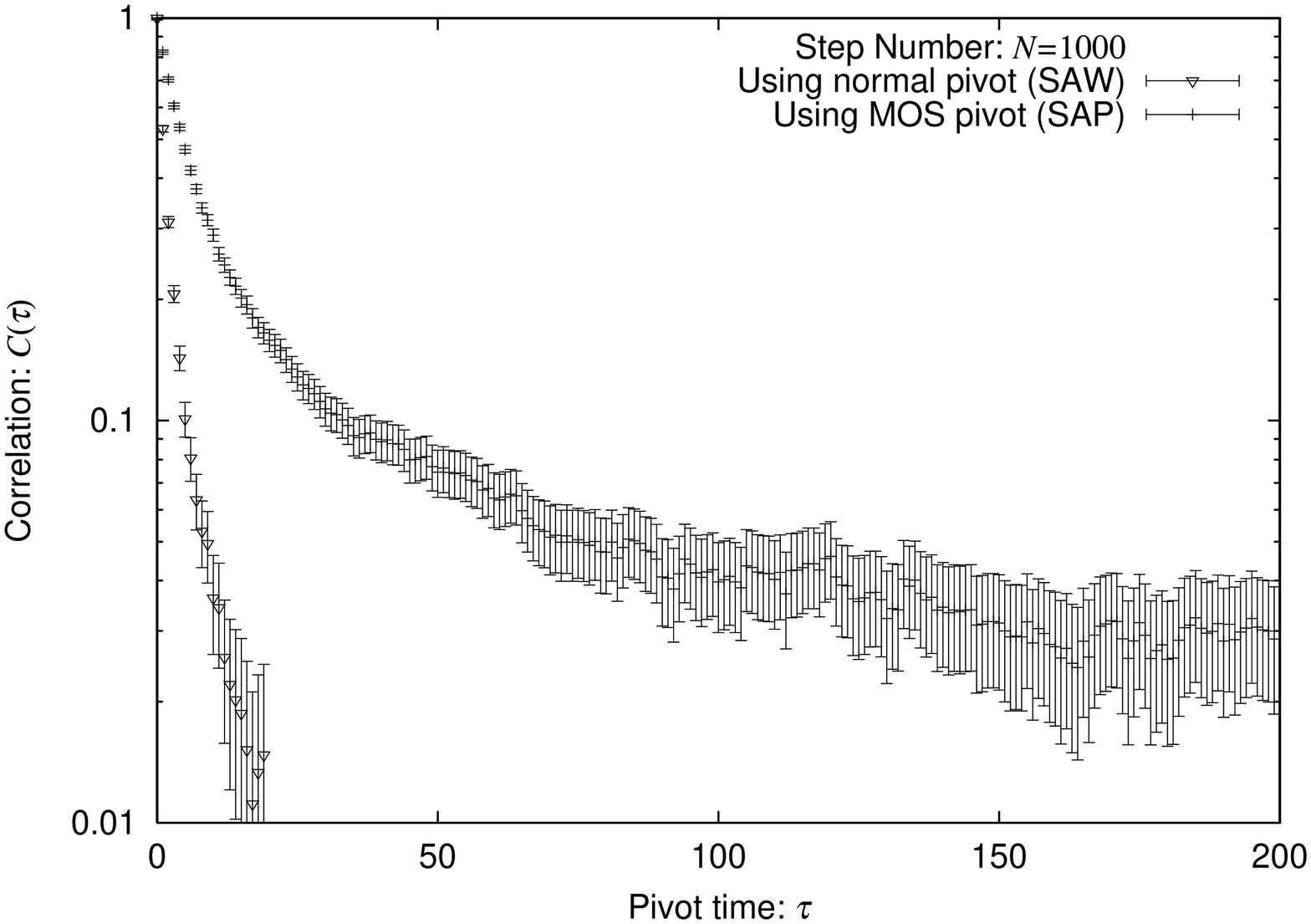}}
\end{minipage}
\end{center}
\caption{Pivot Correlation $C(\tau)$ at $N=1000$}
{
Similarly to the Fig.\ref{correlation1}
we plot the pivot correlations $C(\tau)$ versus the time $\tau$.
The MOS pivot and the normal pivot operations in the case of $N=1000$, respectively.
Here, the $\tau$ is the number of trials, and
error bars correspond to one standard deviation.
At $\tau=200$, we consider correlations $C(\tau)$ as zero focusing on the MOS pivot correlations for $N=500$ and $1000$.
}
\label{correlation2}
\end{figure}